\def\real{{\tt I\kern-.2em{R}}}
\def\nat{{\tt I\kern-.2em{N}}}
\def\snat{{\rm I\kern-.2em{N}}}

\def\realp#1{{\tt I\kern-.2em{R}}^#1}
\def\natp#1{{\tt I\kern-.2em{N}}^#1}
\def\hyper#1{\,^*\kern-.2em{#1}}
\def\hy#1{\,^*\kern-.2em{#1}}

\def\hyperrealp#1{{\tt ^*{I\kern-.2em{R}}}^#1} 

\def\hypernatp#1{{{^*{{\tt I\kern-.2em{N}}}}}^#1}

\def\leaderfill{\leaders\hbox to 1em{\hss.\hss}\hfill}
\def\srealp#1{{\rm I\kern-.2em{R}}^#1}

\def\pars{\par\smallskip}
\def\parm{\par\medskip}

\def\ref#1{$^{#1}$}

\def\m@th{\mathsurround=0pt}
\def\rightarrowfill{$\m@th \mathord- \mkern-6mu \cleaders\hbox{$\mkern-2mu 
\mathord- \mkern-2mu$}\hfil \mkern-6mu \mathord\rightarrow$}
\def\leftarrowfill{$\mathord\leftarrow
\mkern -6mu \m@th \mathord- \mkern-6mu \cleaders\hbox{$\mkern-2mu 
\mathord- \mkern-2mu$}\hfil $}
\def\noarrowfill{$\m@th \mathord- \mkern-6mu \cleaders\hbox{$\mkern-2mu 
\mathord- \mkern-2mu$}\hfil$}
\def\orgate{$\bigcirc \kern-.80em \lor$}
\def\andgate{$\bigcirc \kern-.80em \land$}
\def\inverter{$\bigcirc \kern-.80em \neg$}
%Archived version
\magnification=\magstep1
\tolerance 10000
\baselineskip  14pt
\hoffset=.25in
\hsize 6 true in
\vsize 8.5 true in
\centerline{\bf Operator Equations, Separation of Variables and}
\centerline{\bf Relativistic Alterations}
\bigskip
\centerline{{\bf Robert A. Herrmann}\footnote*{Partially funded by a grant 
from the United States Naval Academy Research Council. }}
\centerline{Mathematics department}
\centerline{United States Naval Academy}
\centerline{572C Holloway Rd.}
\centerline{Annapolis, MD 21402-5002}
\vskip 14pt
\noindent ABSTRACT. In this paper, a simple and  unified method is 
developed that 
predicts the relativistic 
alterations of 
physical measures when the behavior of a  natural system is characterized by 
means of a specific operator equation. Separation of variables is the 
simple underlying procedure. 
\par\bigskip
\noindent Key Words and Phrases. Gravitational redshift, transverse Doppler 
effect, mass alteration, general and special relativity, separation of 
variables, partial differential equations.\par
\noindent 1992 Mathematics Subject Classifications. 83A05, 83C99, 35A08\par
\bigskip
\noindent {\bf 1. Introduction.}  \par
It is often a difficult problem to find a simple and unified mathematical 
approach to physical science derivations.  The term  
``derivation'' refers to the somewhat informal ``proof'' method used within 
the physical sciences that might be formalized when physical axioms are specifically 
included. Often the 
relativistic alterations in physical measures are obtained by means of 
essentially different derivations that are somewhat ad hoc in character and 
may appear to have no simple underlying approach. 
With respect to natural system behavior that can be characterized by means of 
a special operator equation, a unifying approach seems to exist. \par 
In a typical undergraduate differential 
equations course, the method of 
separation of variables is introduced in a first attempt to solve the one 
dimensional heat or wave partial differential equation. This same approach may 
be the unifying factor that allows one to derive the relativistic alterations 
for all natural system measures that are modeled by a specific operator equation. 
 Further, from the mathematicians point of view, a 
derivation in generalized form would be the most appropriate.\par
First, consider the Schwarzschild metric
$$dS^2 = \lambda(cdt^m)^2 - (1/\lambda)(dR^{m})^2 - 
(R^{m})^2(\sin^2\theta^{m}(d\phi^{m})^2 +
(d\theta^{m})^2),\eqno (1.1)$$
where, as usual, $\lambda = (1 - 2GM/(c^2R^m)),$ $G$ is the 
gravitational constant, $M$ the mass of a spherically symmetric homogeneous 
object, $R^m$ the radial distance from the center of the object and 
the superscript $m$ indicates measurements taken for the behavior of a natural 
system influenced by the gravitational field. 
 Next consider  
the basic chronotopic interval (i.e. line element)
$$dS^2= (cdt^s)^2 - (dr^s)^2, \eqno (1.2)$$  
where $c$ is the velocity of light and $(dr^s)^2 = (dx^s)^2 +(dy^s)^2 + 
(dz^s)^2.$ Transforming (1.2) into spherical 
coordinates yields
$$dS^2 = (cdt^s)^2 - (dR^{s})^2 - 
(R^{s})^2(\sin^2\theta^{s}(d\phi^{s})^2 +
(d\theta^{s})^2).\eqno (1.3)$$\par
The subscripts or superscripts $s$
represent local measurements at a spatial point where gravity affects the measurements. The subscripts or superscripts $m$  
represent local measurements, where gravitational effects vanish. These measurements are compared. 
However, expressions comparing gravitational effects at two spatial points within a gravitational field can  be obtained immediately by comparing the effects at each point with the m-measurements where the effects vanish. Further, when astronomical and atomic distances are compared, then (1.3) can 
be assumed to apply approximately to many observers within the universe. 
This is especially the case if an observer is affected by a second  
gravitational field, in which case  (1.3) is used as a local 
line element relating measures for laboratory standards.\par

Following the usual practice for radiation purposes, 
the representative  atomic 
systems are considered as momentarily at rest. Hence
$dR^m= d\phi^m= d\theta^m = dR^s=d\phi^s= d\theta^s =0.$ 
This yields from (1.1) $dS^2 = \lambda (cdt^m)^2$ and from
(1.3) $dS^2= 
(cdt^s)^2.$ Thus, for this atomic system case, the differentials $dt^m$ 
and $dt^s$ are related by the expression  
$$\gamma dt^m=dt^s,\eqno (1.4)$$                   
where $\gamma=\sqrt \lambda.$ \par
\medskip
\noindent {\bf 2. The Derivation Method.} \pars
Suppose that certain aspects of a natural system's behavior are 
governed by a function
$T(x_1,x_2,\ldots,x_n,t)$ 
that satisfies an 
expression $D(T) = k(\partial T/\partial t),$ where $D$ is a (functional) 
separating operator and $k$ is a universal constant. 
In solving such expressions, 
the function $T$ is often considered as 
separable and $D$ is the identity on temporal functions. In this case,
let $T(x_1,x_2,\ldots,x_n,t) = 
h(x_1,x_2,\ldots,x_n)f(t).$ 
Then  $D(T)(x_1,x_2,\ldots,x_n,t) = (D(h)(x_1,x_2,\ldots,x_n))f(t) = 
(kh(x_1,x_2,\ldots,x_n))(df/dt)$ and is an  invariant separated 
form.  
\par
Let $(x_1^{s},x_2^{s},\ldots,x_n^{s},t^{s})$ corresponded to measurements 
taken of the behavior of a natural system that is influenced only by (1.3) and using 
identical modes of measurement let 
$(x_1^{m},x_2^{m},\ldots,x_n^{m},t^{m})$ corresponded to measurements 
taken of the behavior of a natural system that is influenced by (1.1). 
[Note that this is a ``measurement'' and not a  ``transformation'' language 
derivation.] 
Now suppose that   $T(x_1^{s},x_2^{s},\ldots,x_n^{s},t^{s}) = 
h(x_1^{s},x_2^{s},\ldots,x_n^{s})f(t^{s}).$ Assume that $T$ is a universal 
function and that separation is an invariant procedure. Hence, let the values
$h(x_1^{s},x_2^{s},\ldots,x_n^{s})= H(x_1^{m},x_2^{m},\ldots,x_n^{m})$ and 
the values $f(t^s) = F(t^m)$ and $T(x_1^{m},x_2^{m},\ldots,x_n^{m},t^m) =
H(x_1^{m},x_2^{m},\ldots,x_n^{m})F(t^m).$ 
One obtains with respect to $s$ by application of the chain rule
$$ 
\lambda^s=\left({{D_s(h)}\over{h}}\right)(x_1^{s},x_2^{s},\ldots,x_n^{s}) 
= k{{1}\over{f(t^{s})}}{{df}\over{dt^{s}}}=
k{{1}\over{F(t^{m})}}{{dF}\over{dt^{m}}}{{dt^{m}}\over{dt^{s}}}.\eqno (2.1)$$  
With respect to $m,$ 
$$\left({{D_m(H)}\over{H}}\right)(x_1^{m},x_2^{m},\ldots,x_n^{m}) 
= k{{1}\over{F(t^{m})}}{{dF}\over{dt^{m}}} 
=\lambda^{m}.\eqno (2.2)$$ 
Consequently, $\gamma \lambda^s = \lambda^m.$\par
 Suppose that $T = \Psi$ is the total wave function, $D$ is the 
operator $\nabla^2 - p,$ where $n = 3,$ the constant $k,$ and function $p$ are those 
associated with the classical time-dependent Schr\"odinger equation 
for an atomic system as it appears in Evans [2, p. 56]. 
It is not assumed that such 
a Schr\"odinger type equation  predicts any other behavior except that it 
reasonably approximates the discrete energy levels 
associated with atomic system radiation 
and that the frequency of such radiation may   
be obtained, at least approximately, from the predicted energy variations.
The eigenvalues 
for this separable solution  correspond to energy levels $E^s$ 
and $E^{m}$ for such a radiating atomic system.  
Thus $\gamma E^s= E^{m}.$  Radiation occurs when there 
is a discrete change in the energy levels. 
This yields $$\gamma\Delta E^s =\Delta E^{m}. \eqno (2.3)$$
Now simply divide (2.3) by the Planck constant and  obtain the basic 
gravitational frequency redshift 
 expression $\gamma\nu^s=  \nu^{m}$ as stated in Bergmann [1, p. 222].\par
Since this actual derivation is slightly generalized, other operator 
expressions 
can be substituted for $D.$ For another prediction, consider substituting 
the Laplacian $\nabla^2$ 
for $D.$ This would yield for an appropriate object 
an alteration due to the gravitational field 
of the usual temperature function obtained when the PDE for internal heat 
transfer is solved. \pars
\noindent {\bf 3. Additional Applications.}\pars
Consider the special theory linear effect line element [3]
$$dS^2 = \lambda(cdt^m)^2 - (1/\lambda)(dr^m)^2, \eqno (3.1)$$
where $\lambda = (1 - v^2/c^2)$ and $v$ is a constant relative velocity.  
Suppose a 
special theory relativistic effect is considered to take place within 
an atomic system 
itself and is assumed to be the same effect whether motion is transverse or 
receding or approaching the observer, then this is modeled with respect to 
special theory effects by letting  
$dr^m = dr^s = 0$ in (1.2) and (3.1). Hence 
(1.4) holds for this physical scenario. Now
the same argument used to obtain the gravitational redshift can be 
applied in order to obtain the relativistic (i.e. transverse Doppler) 
redshift prediction $\gamma\nu^s=  \nu^{m}.$ Ives and Stilwell [4] were the 
first to experimentally verify this prediction. 
\par
Finally, consider a freely moving particle of mass $M$ moving in a  
``straight'' line with constant relative velocity $v_E.$ 
For a Hamilton characteristic function
$S^\prime,$ the classical Hamilton-Jacobi equation becomes $(\partial 
S^\prime/\partial r)^2 = -
2M(\partial S^\prime/\partial t)$ [5, p. 451]. Suppose that $S^\prime(r,t)
= h(r)f(t).$ Again consider line elements (1.2) and (3.1) while letting the    
universal nature of $S^\prime$ and invariance of separation imply that 
$h(r^s) = H(r^m),\ f(t^s) = F(t^m).$ The same argument used for the relativistic redshift 
derivation again yields equation (1.4).  
Let $D= (\partial(\cdot)/\partial r)^2.$ 
The same procedure used to obtain (2.1) and (2.2) yields
$$ 
\left({{\partial h(r^s)}\over{\partial 
r^s}}\right)^2\left({{1}\over{h(r^s)}}\right) 
= -2{{M^s}\over{f^2(t^{s})}}{{df}\over{dt^{s}}}= M^s\lambda^s_1=$$
$$-2{{M^s}\over{F^2(t^{m})}}{{dF}\over{dt^{m}}}{{dt^{m}}\over{dt^{s}}}
=M^s\lambda_1^m/\gamma.\eqno (3.2)$$  
With respect to $m,$ 
$$\left({{\partial H(r^m)}\over{\partial 
r^m}}\right)^2\left({{1}\over{H(r^m)}}\right) 
= -2{{M^m}\over{F^2(t^{m})}}{{dF}\over{dt^{m}}}=M^m\lambda_1^m.\eqno (3.3)$$ 
In (3.2) and (3.3), the quantities $M^s$ and $M^m$ are obtained by means of  
an identical mode of measurement that 
characterizes   
``mass.'' Assuming that the two separated forms in (3.2) and (3.3) 
are invariant, leads to the special theory mass expression 
$M^m = (1/\gamma)M^s.$
These 
examples amply demonstrate the utility of the separation of variables approach 
in obtaining  
various 
relativistic alterations in measured physical quantities. \parm
\centerline{\bf References.} \pars
\noindent 1. Bergmann, G. Introduction to the Theory of Relativity,
Dover, 1976.\par
\noindent 2. Evans, R. D. The Atomic Nucleus, McGraw-Hill, 1955.\par 
\noindent 3. Herrmann, R. A. An operator equation and relativistic alterations in the time for radioactive decay, {\sl Internat. J. Math. Math. Sci.} (To appear)\par  
\noindent 4. Ives, H. and G. Stilwell. Experimental study of the rate of a 
moving atomic clock, {\sl J. Opt. Soc. Am.,} {\bf 28}(1939) 215--301.\par                                               
\noindent 5. Synge, J. L. and B.A. Griffith,  Principles of 
Mechanics, McGraw-Hill, 1959.\par 
\end